\documentclass[fleqn,11pt]{wlscirep}
\usepackage[utf8]{inputenc}
\usepackage[T1]{fontenc}
\usepackage{bm}
\usepackage{wrapfig}
\usepackage{framed}
\title{How to set up your first machine learning project in astronomy}

\author[1,*]{{Johannes} {Buchner}}

\author[2]{{Sotiria} {Fotopoulou}}

\affil[1]{Max Planck Institute for Extraterrestrial Physics, Giessenbachstrasse, Garching, 85748, Germany}

\affil[2]{H.H. Wills Physics Laboratory, University of Bristol, Tyndall Avenue, Bristol, BS8 1TL, United Kingdom}
\affil[*]{e-mail: johannes.buchner.acad@gmx.com}

\begin{abstract}
Large, freely available, well-maintained data sets have made astronomy a popular playground for machine learning projects. Nevertheless, robust insights gained to both machine learning and physics could be improved by clarity in problem definition and establishing workflows that critically verify, characterize and calibrate machine learning models. We provide a collection of guidelines to setting up machine learning projects to make them likely useful for science, less frustrating and time-intensive for the scientist and their computers, and more likely to lead to robust insights. We draw examples and experience from astronomy, but the advice is potentially applicable to other areas in science.
\end{abstract}
\begin{document} 
 \flushbottom
\maketitle

\thispagestyle{empty}

\noindent \textbf{Key points:} 
\begin{itemize}
    \item A machine-learning project should have a specific, measurable and useful objective, that can be validated with a test harness. 
    \item It is good practice to establish a set of trivial baselines, and to define what a significant improvement over those would be.
    \item For scientific applications, ingesting and reporting uncertainties is expected.
    \item Ablation studies and generative component building can lead to interpretable insights.
    \item To learn which machine learning method is likely to be successful requires publishing successes and failures of systematic tests.
\end{itemize}

\vfill

\noindent\fbox{
\makebox[\textwidth][c]{\begin{minipage}{\textwidth}\centering
\vspace{1cm}
\Large{
Please find the more polished, published version at:

\url{https://www.nature.com/articles/s42254-024-00743-y}

Full-text access: \url{https://rdcu.be/dQl5O}}
\vspace{1cm}
\end{minipage}}}

\clearpage

\begin{framed}
\section*{Dictionary}
The dictionary below is intended to help communication between people with backgrounds in astronomy, computer science, and statistics.
It translates machine learning terms (left) for astronomers (right).
Beware that astronomy jargon already includes precision, power, intensity, volume, feature and target with different meanings than in machine learning. 
Because of the ambiguity, it is best to avoid these or use them with caution and clarity.
The following explanations might help.
Related to the input data: \\
\textbf{(row) observation, instance, point}: object, sample, catalog entry \\
\textbf{(column) input features}: data, such as catalog columns or pixels, used for fitting a model \\
\textbf{label}: ground truth value that should be predicted, for example star or galaxy class (categorical), or redshift (continuous). In supervised ML, this drives the training, in both supervised and unsupervised ML it is used to quantify the performance of the model. \\
\textbf{target (supervised ML)}: model prediction \\
\textbf{pairs plot}: corner plot, plot each pair of catalog columns against each other. \\
Related to training: \\
\textbf{ML algorithm}: method, such as a specific neural network architecture with a chosen loss function and training procedure, or a random forest with chosen hyper-parameters. \\
\textbf{model}: model with all of its parameters fixed to the best fit, capable of prediction. \\
\textbf{training}, learning: fitting a model. In a neural network, this includes the free parameters, the linking structure and functions, the noisy minimiser that finds generalizing solutions \cite{Mandt2017} and how it is steered. \\
\textbf{cross-entropy loss}: for categorical data, information loss between data and model \\
\textbf{L2 loss}: chi$^2$ statistics; assuming a Gaussian distribution for the difference between data and model. \\
\textbf{Gaussian negative log-likelihood loss}: Model predicts mean and standard deviation of a Gaussian for each data point \\
\textbf{scoring function}: quantifies how far the predicted value is to the truth; loss function, fitness function, minimization function \\
\textbf{scoring rule}: same as scoring function, but for models that predict probability distributions \\
\textbf{proper scoring rule}: The score is lowest when the truth is predicted. \\
Related to model performance: \\
\textbf{unbiased (regression)}: When averaged over infinitely many samples, the difference between truth and prediction is centered at zero. A model may be unbiased on on the training sample, but biased on another sample. \\
\textbf{consistent}: In a single example, if more or better data is ingested, the prediction converges to the truth. \\
\textbf{estimator variance}: square of the standard deviation of the difference between truth and prediction. \\
\textbf{sensitivity, recall, power}: true positive rate, completeness, e.g. how often the detection triggered on a real signal. \\
\textbf{precision}: 1 - false positive rate.
\end{framed}

\section{Introduction}

Today, no astronomy conference is complete without machine learning (ML)
sessions. Examples of successful applications include the discovery of supernovas, image analysis of galaxy morphology, and removal of contaminating emissions in various data pipelines.
Undoubtedly, there is excitement about these methods. It can also not be neglected that many science graduates and undergraduates transition from a science-focused ML research project into data science careers in information technology companies, which drive ML research.

Consequently, research institutes across the world have started applying
machine learning methods developed in industry for astronomical data
sets, in the hope of accelerating tedious tasks or data exploration
towards new insights. 
Collaborations between data scientists and astronomers have become more common, with researchers with a background in astronomy trying to become more knowledgeable in ML, or computer scientists diving into astronomy data sets.
This situation is hardly unique to astronomy. See for example, \cite{shearer2000crisp,saltz2015need,Martinez2021,Artrith2021chemistryMLbestpractices}. 

This recommendation document addresses the need for specific guidelines. Intentionally, specific ML methods and models are excluded from the discussion here. See \cite{garofalo2016astrophysics,Zhang2015,Lahav2023} for the characteristics of large astronomical data sets, \cite{Borne2009,Djorgovski2022AIreview} for brief general overviews, \cite{Fluke2020} for a review of the ML maturity in different areas of astronomy, and \cite{Ivezic2019,Baron2019MLbasics} for a hands-on guide through techniques \cite{hackeling2017}, and dedicated reviews of classification \cite{Graham2017,Yang2023}, unsupervised learning (Fotopoulou, sub.), clustering \cite{Yang2022clustering} and active learning \cite{Settles2009,Lochner2021}.

The intended audience of these guidelines is students or supervisors of students, practicing ML with astronomical data sets.  
It is assumed you are familiar with ML fundamentals from generic tutorials. 
For example, know the difference between unsupervised and supervised learning and 
how they are used, know the concepts of regression and classification, 
be comfortable with pre-processing steps such as data normalisation and 
exploration, have applied some machine learning methods to a benchmark 
dataset such as MNIST \cite{deng2012mnist}, and be familiar with common evaluation metrics 
such as false/true positive rate, precision, recall (or equivalently, 
purity and completeness), with commonly used loss functions, and with standard practices such as cross-validation.
For non-astronomers, the necessary background to understand illustrative examples is provided.

The insights provided are intended to help
(1) reduce the number of months spent on projects, 
(2) focus computing power on tests that lead to helpful insights,
(3) guide towards a useful end product that advances science,
(4) produce insightful and impactful papers and
(5) improve the career development of the machine learning practitioner by adopting best practices.
We hope that this guide will lead ML practitioners to understand the challenges of developing an ML pipeline in astronomy, and astronomers gain insights on how to approach, build, evaluate and improve machine learning models.

\section{Your new project}

\subsection{Have an objective}

This section discusses what to do before you actually do machine learning.
A project is an activity for a limited time. How do you limit the
time? How do you know the project is \textbf{complete} or has been a success?


Start by defining the problem. A problem statement clearly lays out
the question you want to answer and the technical criteria the software
needs to accomplish. Common advice is to have SMART objectives \cite{doran1981SMART}: Specific, Measurable, Achievable, Relevant, and Time-Bound.

An example may help clarify what we mean. A poor problem statement is 
``Out of this pile of images, I want to identify pairs of galaxies that are in the process of colliding''. 
Why? If I gave you a tool that can do it poorly, you would be unsatisfied. You would not be done with the project, for a reason that is not specified in the problem statement. Therefore, the problem statement is incomplete. \cite{saltz2015need} discusses this issue from a data science methodology perspective.

\subsection{Have a good objective}

Let's see what makes a better problem statement, through an analogy.
A medical doctor might choose to only perform a test when the outcome leads
to different consequences. If they have the same treatment for both
possible diseases, they could proceed to the treatment. By analogy,
think about what classes you care about, and how the next step drives these
definitions.

A better problem statement is: ``It is unknown whether merging galaxies 
make up 1\% or 0.01\% of this sample of 10,000 galaxies. 
I want to develop a method that can help me distinguish
between these two hypotheses.'' This is better because it tells us
something about the quality of the tool we need. If I told you the
true positive rate (completeness) and false positive rate (purity) of my method, you would be able
to assess whether it will be able to achieve your objective. Statisticians
call this power analysis, and it is mandatory in government-funded medical studies before taking data \cite{bausell2002power}.

Another good problem statement is: ``We need to know whether transients
are class A or B or C, because for B and C we will immediately 
point a telescope with a spectrograph at it for 1 hour, 
whereas for class A, we will take a series of brief images over the 
next ten days.'' Here you have some investment
why it matters. Flesh this problem statement out with the cost of
one false classification, and then you can derive the needed false
positive and true positive rates. 
From the problem
statement, since the follow-up
action does not need to distinguish between B and C, only a A vs BC classifier is needed.
While an A vs B vs C classifier may perform well on an A vs BC task, this is not guaranteed. The performance metric may be wrongly chosen to not optimize the model for A vs BC confusion, and give too much importance to B vs C confusion. Related, the number of members in the classes (class imbalance) may need to be considered when designing an objective that minimizes confusion.

Think about the \textbf{operational definition} of your classes. 
Within the context of the study, operational definitions \cite{operationaldefinition} are introduced to define how the classes are connected to observational data of the experiment.
Take as an example the task of classifying sources as stars, galaxies or Active Galactic Nuclei (AGN). Stars and galaxies are mutually exclusive classes, but all galaxies have nuclei that could be active to some level. So AGN luminosity is a continuous attribute of galaxies. This would 
more naturally fit a regression paradigm, where AGN luminosity is a parameter of interest, than classification. If we can still set this up 
as a classification problem, operationally this then corresponds to 
placing a cut on AGN luminosity, which separates the galaxy population
into a galaxy class and an AGN class.
You should understand how that cut behaves (for example, how it depends 
on the luminosity of the galaxy's stars, or the type of AGN), because
that is part of your operational definition.

Historically, astronomers have classified new phenomena by their observational
appearance first. As understanding progresses, the observational classes are associated with astrophysical mechanisms. Clarify first which of the two you need: 
an observational classification or classification by physical process?
It may be easier to classify by observational facts if they are potentially present in the training data
(for example, from a single spectrum a hydrogen-poor and a hydrogen-rich supernova can be distinguished, 
while the physical explosion mechanism may not be identifiable \cite{Minkowski1941SNclasses}). This also applies to regression. To give
one example, you could build a predictor of the correction factor to convert flux to luminosity
(essentially, luminosity distance), but if the input data is spectroscopy, in actuality the data 
is tracing the location shift of spectral features (redshift).
Having a well-defined, smaller problem should be easier to address
and may also be more reusable.

Formulate your problem statement as precisely as possible. Include two
scientific scenarios you want to distinguish, or technical requirements
the method has to fulfill.

\subsection{Define a finish line}

Machine learning is optimizing a performance measure, typically minimizing a loss function. Only if this measure
is meaningful, and if we know what threshold we need to get to, we can declare the project finished.

Once you have an objective written down, design a performance measure 
that quantifies how close models come to reaching the objective. 
This will be specific to your question (more on this below),
and may be a custom score.

Ask yourself: \textbf{what performance would be good enough to close
the project?} For example: A classifier with 98 percent true positives and 1 percent false positives would be a meaningful improvement over the current workflow.
To how many digits is the loss function performance meaningful before other problems dominate? 
For example, weak systematic issues may come in when applying the method in the real world, or the size of the test data set allows estimating the rates to an accuracy of 0.5 percent. So two ML models that give 98.1 and 97.9 percent may both reasonably fulfill the target of 98 percent within the limitations of the test.

Once that level is reached, you may want to focus
attention on the next biggest source of error toward your science
goal. Below are some further examples for judging supervised and unsupervised learning.

\subsubsection{Comparing classifiers}
Let's consider classification. 
How do you know one classifier is better than another?

First, derive from your objective whether you want \textbf{high completeness
or high purity}, or something else. High completeness is required when exploring a rare
class of objects which should receive further observations, e.g. supernova. High purity
may be required for discovery claims or sample studies, e.g. galaxies.
Between two classifiers, the one with the better performance measure is preferable.

If you have two groups of scientists, one interested in class A, and
one in class B, you will need at least two different selection methods:
Some A-scientist will want a pure sample of As, some B-scientist
will want a pure sample of Bs, and yet others may want complete samples. For example, in a typical classification problem of star/galaxy separation, the star classifier might be needed for modelling the response of the instrument (A), while the output of the galaxy classifier will be used for further population studies (B).

\begin{figure}
\begin{centering}
\includegraphics[width=\textwidth,trim=2cm 3cm 1cm 3cm,clip]{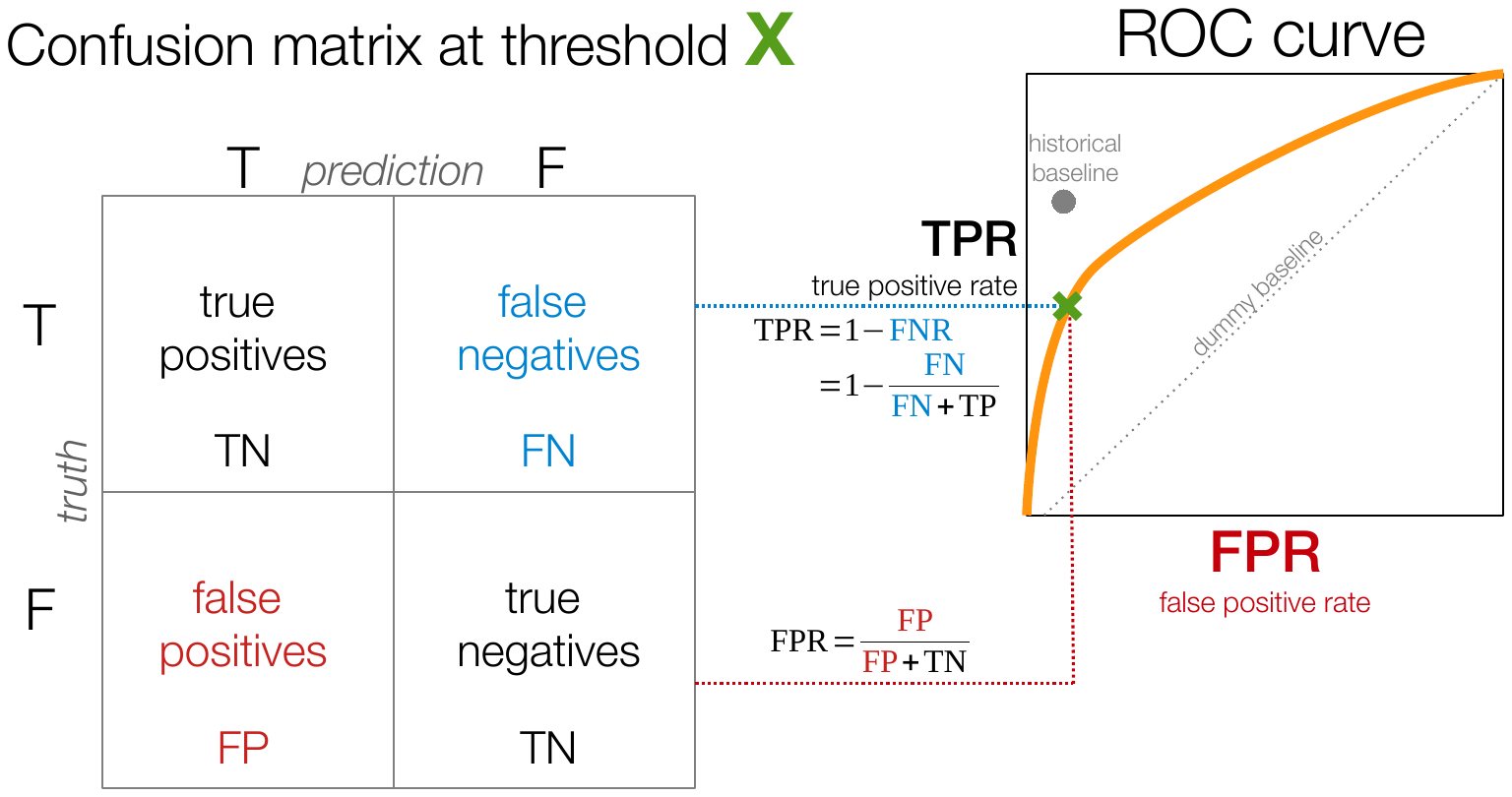}
\par\end{centering}
\caption{\label{fig:roc}The confusion matrix (left) reports the classification numbers needed to compute the false positive rate (FPR) and the false negative rate (FNR) of a classifier at a certain threshold X, for example the classification probability threshold. 
If computed with k-fold cross-validation, these rates can be reported with uncertainties.
Any confusion matrix corresponds to a position on the receiver operator curve (ROC), and varying the threshold gives a curve (orange). A dummy baseline and a historical baseline are included for comparison (see section \ref{sec:baselines}). For choosing among methods with a desired false positive rate, one may calibrate the threshold X, then pick the one with the lowest false negative rate.}
\end{figure}

A sensible model comparison is: With both methods calibrated to 1\% false
positive rate, which has the better true positive rate? This is a
slice through the Receiver operator characteristic (ROC) curve, but encodes what we want to use the method
for (pure samples, in this case). An alternative for high completeness
would be: ``With both methods calibrated to an 1\% false negative
rate, which has the better false positive rate?''
The relation between false positives, false negatives and the ROC curve is illustrated in Figure~\ref{fig:roc}.

\subsubsection{Comparing regressors}

For regression, similar questions arise. For example, when minimizing
a Negative Log Likelihood of a Gaussian (L2 loss) and considering
the performance over the test sample, do you care about the average
error or the worst case? Perhaps you want to consider the maximum
absolute deviation to be less sensitive to outliers? This depends
on the problem statement.
Is an equally weighted L2 loss the statistic that measure the performance needed to achieve the stated objective, or should a custom performance measure be crafted?

\subsubsection{Comparing unsupervised methods}

Unsupervised learning is one of the most challenging to judge. Your
enemies here are (1) the human brain's tendency
to see patterns in unclear outcomes, also known as \textbf{pareidolia}, and 
(2) \textbf{wishful thinking} that the project is successful, that the machine is smart
and insightful, and that your group can go out and proudly show off
a shiny new tool. Be skeptical. Devise tests for \textbf{external
validity}. For example, get independent data upon which something
meaningful was predicted.

To give an example: Hierarchical clustering algorithms have a threshold parameter
that decides when to break up clusters. If the number of clusters
created is arbitrarily modulated with that threshold, then the number
of clusters produced cannot be the sole criterion for preferring one
method over another.
As another example, k-means clustering can only produce convex (ellipsoidal) clusters, with a pre-determined number of clusters. 
So it may be helpful to test whether the 
hyperparameter choices have been appropriately set or explored. 
Ideally, the chosen approach should match the problem characteristics.

Dimensionality reduction techniques are not entirely without guidance.
A crucial input is the metric of similarity between two observations, 
which should be chosen carefully. Recall that a simple 
dimensionality reduction is to throw away all but two features and 
plot them against each other. How can you characterize whether another 
dimensionality reduction technique is better?
A semi-supervised technique would take some labeled examples and test whether the same labels co-occupy the same regions in the reduced space. This can help judge, quantitatively or qualitatively, whether one technique is better than another.

\subsection{Create a test harness}

To be sure a student learned something successfully, schools prepare 
test exercises the students have not seen before. Carefully crafted exam
questions can validate important abstract concepts. This is also
a programming technique: To be sure a function is correct, 
\emph{unit tests} verify whether given a certain input, the
function output is as expected.

Combining these two ideas, you could create a test harness which
the method, if it really understood something, should easily
pass. If the method is ultimately trained on real data, 
can you first test the pipeline on data generated from a high-quality simulation,
to understand its biases?
Furthermore, perhaps a small, high-quality observation data set is 
available as independent ground truth to cross-check the model.
For imaging, perhaps a sane outcome can be expected for certain hand-crafted shapes 
(such as a two-dimensional Gaussian on a plateau of noise).

You can then also come up with some more adversarial examples. How
sensitive is it to variations? Can you perturb training samples
by deleting or adding features, that the method should be invariant
to? 
A more general approach is to combine two observations into a new hybrid.
For example, adding background noise from one observation to another should give the same outcome. The sum of light from two galaxies should still be recognizable
as a galaxy (as a chance alignment or simplistic galaxy merger).

\section{Work environment}

In this section, we discuss some non-technical aspects of approaching
machine learning problems.

\subsection{Understand your role}

Will machines eventually replace data scientists? Not 
necessarily! Machines are very good at iterating
through a more or less random selection of methods, tweaking 
parameters, and returning the best performing method \cite{ThorntonAutoWEKA,Erickson2020AutoGluon}.
To beat the machine, do something machines cannot do. 
You can create success consistently if you train to (1) understand the 
specific problem in depth and (2) understand the context of the work and its 
interface with humans. This is expanded below.

\subsubsection{Interface with the domain experts}

A major component of computer science is becoming familiar with the
problem domain, and understanding and specifying with the end-user the
problem statement. Become good at that process. It will be a strength
for your career, and a differentiation capability. This involves being
active in solving the problem and asking many questions.

If you are not an astronomer, ask your astronomer colleagues many questions.
They will be more than happy to explain. 
Astronomers have developed lots of jargon that may be
difficult to pierce through for newcomers. If you are lucky, you have a multi-lingual
astronomer who knows which field you come from and what you know and
can explain clearly without resorting to such short-cut terminology.
Otherwise, you may need to become familiar with the terms,
by asking repeatedly for explanations and definitions.

Iterate in the team until the objective is clearly 
defined and written down (see section §2). Build up your understanding of the meaning
of input features by exploratory data analysis techniques. For example,
make a pair-wise input feature scatter plots (pairs or corner plot), colour-coded by target
label or value. With domain experts, try to understand all the apparent trends and features.
Are there missing data, and how should they be treated?

\subsubsection{Open the black box}
\label{subsubsec:blackbox}
Machine learning models can be abstractly understood as input-output-mapping black
boxes, which were optimized against some loss function. Do not stay on this
level. Go deeper, and understand the underlying reason why the model
or data behave this way, or at least characterize the essence of its behaviour. 
To improve your skills as a data scientist, you may need to explore the 
process that created the training data, adjust the model manually, and 
\textbf{use plenty of visualizations}.

There are probably imperfections in the input and output features you have received. 
To address this, your team needs to be able to interface backward to the
data producer, or apply modifications. Therefore, for each
input and output feature, understand the process by
which that feature has been produced, at least to some level. 
The steps above require effort and may be intimidating, but is a productive and professional approach.
It will inform you what to spend time on optimizing, and set you apart from less experienced data scientists.

For example, astronomical catalogs are generated from images, by detecting 
light sources. In ML parlance, this in itself is already a feature-extraction 
process. During this process, a number of choices are made, such as:
\textit{Should the detection be optimized for completeness or purity?} 
\textit{How many pixels contribute light to a source?} 
\textit{How do I define the boundary of one galaxy?} The choices will 
differ for teams looking for supernovae, and teams studying galaxy evolution 
across cosmic time, and a general-purpose solution is difficult \cite{Lieu2019}. 
Quality flags are typically given alongside the detected sources. 
These need to be understood and considered.

Gather insights about the performance of a trained model. For example, when 
you get misclassified outputs, compare the inputs of these samples to 
the correctly classified cases. Ask how these input data came to be 
different, and which features may make them outliers. This may give 
you ideas on additional constraints (such as 
symmetries) you can place on the model and assumptions you are allowed 
to make (or not). Perhaps additional information not included 
in the model can be retrieved to help understand artifacts.
Even if no changes are made, this may help you build an intuition
of the capabilities and limitations of the method.

Techniques for investigating the internals of models have been developed. These include interpretable models (e.g., decision trees), interpretation techniques that are model-agnostic (e.g., surrogate modeling) and techniques specific to neural networks (e.g., saliency maps). See for example \cite{molnar2022,rudin2022interpretable} for recent reviews.

\subsection{The person and the method}

As you or your student spend time on a project, there is a tendency
to entangle the person with the method. For example, ``My student
cannot classify type 1 supernovas well``, or ``I achieve
80\% precision on this data set''. Beware however, that the person,
the quality of their work and the performance of a model are three
separate aspects. The quality of work building and characterising
a method can be excellent, yet the model may have some intrinsic limitations.
For example, the training data may not contain the 
information to achieve the desired outcome, to no fault of the person.
Thus model performance does not necessarily increase with amount and quality
of work.

Upon publication, it is crucial to give credit, such as by authorship, to those who contributed to a successful ML project. This includes the data scientist but also data providers (such as surveys), data engineers who cleaned the data and defined training samples, programmers, software toolkits, computing infrastructure, etc.

\section{The starting point}\label{sec:baselines}

By now you have an objective and a performance measure, and are motivated
to do excellent work. But wait a minute before applying the glorious
machine learning technique and heating up the data center.

\subsection{Establish a `dummy' baseline}

First, understand the performance of the simplest method.
For classifiers,  return the most common class, for regression, return
the median training sample output. The `dummy'\footnote{Module \url{https://scikit-learn.org/stable/modules/classes.html\#module-sklearn.dummy}, discussed in \url{https://scikit-learn.org/stable/modules/model_evaluation.html\#dummy-estimators}} module in scikit-learn
has many more variants. As an illustrative example, if your data set
is 90\% class A, then a dummy classifier will get it correct 90\%
of the time. That should be the dummy baseline. Include it in your
performance measures and plots (see Figure~\ref{fig:roc}).

For unsupervised learning, a simple dimension reduction technique
from N dimensions to two is to drop all but two features. This will
still give you a space, and raise the question: how do you know a
fancier technique is better than this dummy technique?

\subsection{Establish a historical baseline}

Typically projects do not start in the void. They try to improve over
some classical method. Include the predictions of that method in 
performance measures and plots. This will help build intuition whether
what you are developing is advantageous.

It may be the case that you do not beat this baseline. And that can
be a success of the project when the objective is to achieve comparable
performance, but 10,000 times faster.

\subsection{Establish an initial baseline}

Add a baseline from a simple or widely-applicable standard method. 
For classification, use logistic regression, and a random forest with default parameters. 
For regression, use linear regression and random forest regression.
The goal here is to see how far you can get before spending any time on tuning.
Well-tested, user-friendly, high-level frameworks like scikit-learn allow 
achieving a result relatively quickly.

The methods mentioned above are suitable for low-dimensional, tabular data.
In the case of high-dimensional data such as images, some 
basic feature engineering may make them usable. For example,
a 100x100 pixel galaxy image can be divided into six radial shells,
and in each annulus the mean and standard deviation of the pixel values
computed. This gives 12 features instead of 10,000. On one or two dozen
input features, a wider range of simple methods is applicable.

Such an initial baseline can help check whether a more elaborate method was
able to extract something non-obvious.

\subsection{Be robust to sample variations}
\label{sec:variations}

Many methods are sensitive to random generator seeds, and even the order and
portions of the training sample used to train them. Therefore, run
several times with different seeds and apply K-fold cross-validation 
to the training/test input, and report in addition to the mean also 
the standard deviation of the performance indicators.
This will include (1) variations in training, (2) sample variance
in the training and test sample, (3) the effect of small number 
statistics, which may limit the precision to which the performance
can be reported. 

Reporting performance quantifiers with error bars allows judging how
trustworthy they are, and how many digits are significant.
When comparing two methods, and these error bars overlap, you can
consider the two approximately equal. This avoids thinking a change 
made a substantial improvement, when it was merely random chance. 
In turn, this will save you from spending time on pursuing statistical flukes.

\subsection{Negligible improvements}
\label{sec:insignificant}
What performance difference between two methods is a \emph{meaningful}
improvement vs. basically equivalent performance? If the classification
accuracy improves by less than 0.1\%, would you really prefer a method that is more complex, slower or more difficult to train?

Define a threshold within which two methods can be considered equal.
If useful, add secondary criteria that break the equivalency, such
as run-time performance. For example, we may prefer smaller and
more explainable methods (linear regression / logistic classifier
> random forests > deep neural networks) given comparable performance.

\section{Being useful for science}

Now that we have our baseline down, let's discuss the scientific needs
of a machine learning method.
In some industry applications, it may be accepted to take a guess, and society is expected to somehow deal with the small number of mistakes (for example, house number recognition for Google Maps).
However, serious discriminatory consequences of biased AI systems have occurred, calling for a more professional approach\cite{mehrabi2021aibiassurvey}.
In science, we often wish to distinguish two
hypotheses from a starting point of not knowing which one is true.
We can have \emph{three} outcomes: (1) A is true and B is not, (2)
B is true and A is not, or (3) we still do not know. Most often, the
outcome may be ``it's more complicated than we thought'', and new
questions arise, which is also a valuable outcome. This not-knowing
and cautious approach is an essential part which should be captured
by the machine learning method. Indeed, the more tests you perform,
the better you can explain or characterize what is going on, and 
the more cautious and conservative you are in your statements, 
the more scientists inside and outside your institute will believe 
in the model.

\subsection{Embrace uncertainty}

In \textbf{physics}, every measurement is accompanied by an uncertainty.
While the measurement device would produce raw data, 
data reduction and calibration annotate these to higher-level
quantities with uncertainties. Three cases are very
common \textbf{in astronomy}:

(1) Measurement with a Gaussian error bar. In this case, you could
augment the data set during training by randomly drawing a Gaussian
random variable with that mean and standard deviation.
Augmenting the data set with random realisations is different in subtle ways than encoding a probability distribution function (PDF). This is because a trained model cannot infer an updated PDF on such inputs, when they are provided as realisations (without uncertainties). Techniques such as Bayesian Neural Networks can ingest PDFs. 

When working with multiple columns as described above assumes the uncertainties are uncorrelated. This may be valid, for example, when they are derived from independently measured data. If this is not the case, correlations may need to be incorporated, with a covariance matrix.
Independent of whether the uncertainties are correlated or uncorrelated, or provided at all, input features may be correlated with each other. Normalising the input features may be beneficial, for example, by projecting to a feature space where the input feature covariance is a diagonal, unit matrix.

(2) Non-detections. These occur when a measurement was taken but
the astrophysical source was too faint. 
Sometimes non-detections are encoded as a special value such as -99, or indicated by a flag column. In either case, the nominal value needs special treatment, different from (1) above. 
A non-detection is information, as it
excludes some possible values. This should be given to you as
an bound on possible values. For example, the flux (or magnitude) limit of the survey depth sets the upper (or lower) bound.
Here you could use a uniform distribution that extends
from a reasonably low value to the upper bound. 
If necessary, ask a colleague with relevant expertise what is reasonable. 
If values are always positive (such as fluxes or cosmological redshifts),
you may want to work in transformed (logarithmic) space. If the measurement process
is more or less uniform you can guess the upper bound from the lowest
sample where (1) is true.

(3) No measurement taken. Because astronomy combines many types of
information over the electromagnetic spectrum and beyond, it may be
that the telescope needed to obtain a piece of information has not
been pointed there at all. In this case, we truly lack information
for this sample. Ideally, we like methods that can take still take
a guess if an input feature is absent. For example, if you are using
a neural network, placing a drop-out layer right
after the input features \cite{Srivastava2014} creates realisations
where features are absent (Singhal, Fotopoulou, et al., in prep).
This allows predictions with and without the feature.

These uncertainties are then propagated through the data analyses. Each higher-level output is then also accompanied by an uncertainty.

\subsection{Produce uncertainties}\label{sec:makeuncertainties}

Hopefully, the ML model may ultimately be embedded in a larger scientific workflow,
where the output prediction is combined with some physical model inference.
In modern astronomy, the Bayesian approach is widespread, where instead
of a single number output (point estimate), one thinks of a probability distribution.
A subsequent analysis that ingests the outputs of the ML model may be capable of handling a full probability distribution. It may even require it.

A first level of producing such probability distributions
is to adopt the global performance quantifiers. 
For example, in regression, the root mean square over the test data can be adopted
as a standard deviation for a Gaussian distribution centred at
the predicted number.
A higher level is to predict probability distributions of 
different widths or shapes for each sample.

\begin{figure}
\begin{centering}
\includegraphics[width=0.7\textwidth]{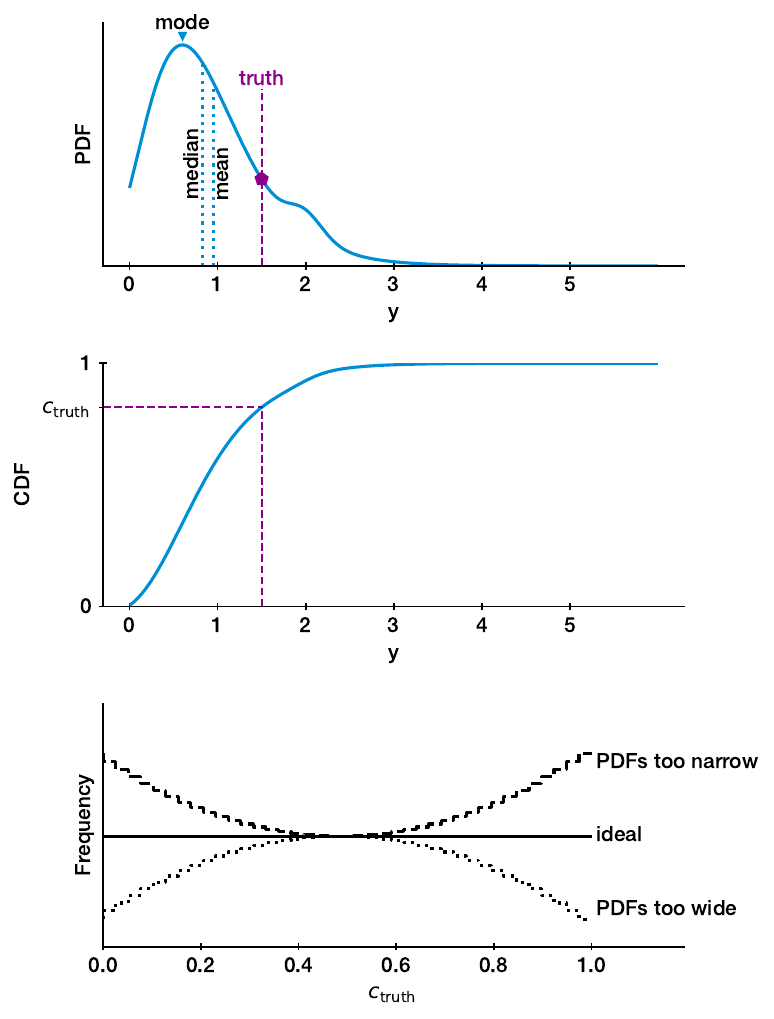}
\par\end{centering}
\caption{\label{fig:ucalib}From the predicted probability density
function (blue curve in the \textit{top panel}) of one validation sample, the probability of the true
value (purple) can be read off ($p_\mathrm{truth}$).
The higher the product of these probabilities over the entire validation sample, 
the more informative the prediction.
The \textit{middle panel} shows the corresponding cumulative probability function,
from which the cumulative probability of the true value can be read off ($c_\mathrm{truth}$).
The histogram of $c_\mathrm{truth}$ values over the validation sample is illustrated in the
\textit{bottom panel}. If the PDFs are overly narrow, the $c_\mathrm{truth}$ values are frequently 
near 1 or 0 (dashed). Ideally, the values show a uniform distribution. Based on \cite{Hamill2001}.
}
\end{figure}

But before going into involved methods, we need to be able to judge 
whether one way of producing uncertainties is substantially better than another.
Figure~\ref{fig:ucalib} provides an illustration of
a predicted PDF and how to characterize 
with the validation sample whether the PDFs are informative 
and well-calibrated.
The information loss of going from the true value
to the predicted PDF is given by the probability density at the true
value (purple pentagon in the top panel of Figure~\ref{fig:ucalib}). 
Because PDFs are normalised, overly wide distributions yield lower probabilities.
Since all validation samples need to be fulfilled simultaneously,
the product of these probabilities provides the recommended performance quantifier.
This corresponds also to the mean log-likelihood loss. 
Higher values indicate more informative methods.
The reliability of the PDFs can be characterized by the location of the true value on the
cumulative density function (middle panel of Figure~\ref{fig:ucalib}).
The histogram of cumulative probability values over the validation sample
 (middle panel of Figure~\ref{fig:ucalib}) can judge whether the PDFs are
overly wide or narrow \cite{Hamill2001}. 
The idea here is to verify the output error bars
really correspond to 68\% in the test/validation data 
(see Table~2 in \cite{Ghosh2022} for an example from galaxy morphology).

Producing uncertainties can be achieved by an ensemble of predictors and 
considering the mean and spread. K-fold cross-validation 
can generate an ensemble of predictors, and crudely summarized 
as a Gaussian based on the observed mean and standard deviation of the predictions.
Error bars can also be produced by predicting mean and standard deviations
assuming Gaussianity (under a Gaussian negative log-likelihood loss, or 
by predicting means first and then predicting the absolute deviation between prediction and truth).
More advanced is to adopt a method capable of predicting a full PDF, such as 
mixture density networks or Bayesian neural networks.

Finally, systematic uncertainties that cannot be addressed
should be discussed in the publication.

\subsection{Beware that methods have priors}
\begin{figure}
    \centering
    \includegraphics[width=0.4\columnwidth]{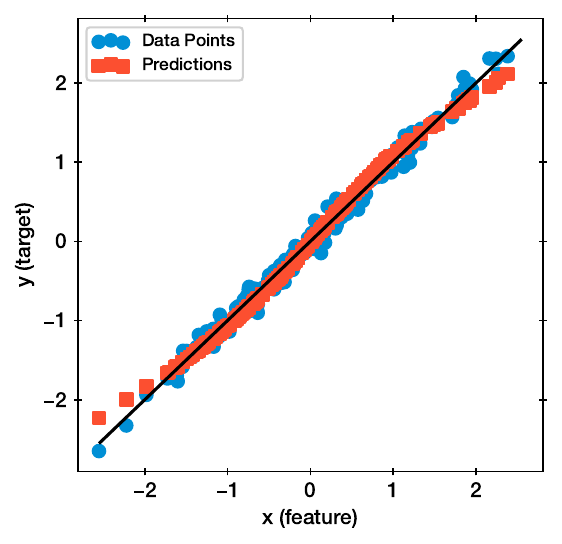}
    \caption{At distribution tails, regression trends towards the bulk of the training sample's target. A normal distributed sample was generated (blue), and K-folded out-of-sample prediction of a neural network is shown (red).}
    \label{fig:regression-bias}
\end{figure}
\begin{figure}
    \centering
    \includegraphics[width=0.7\columnwidth]{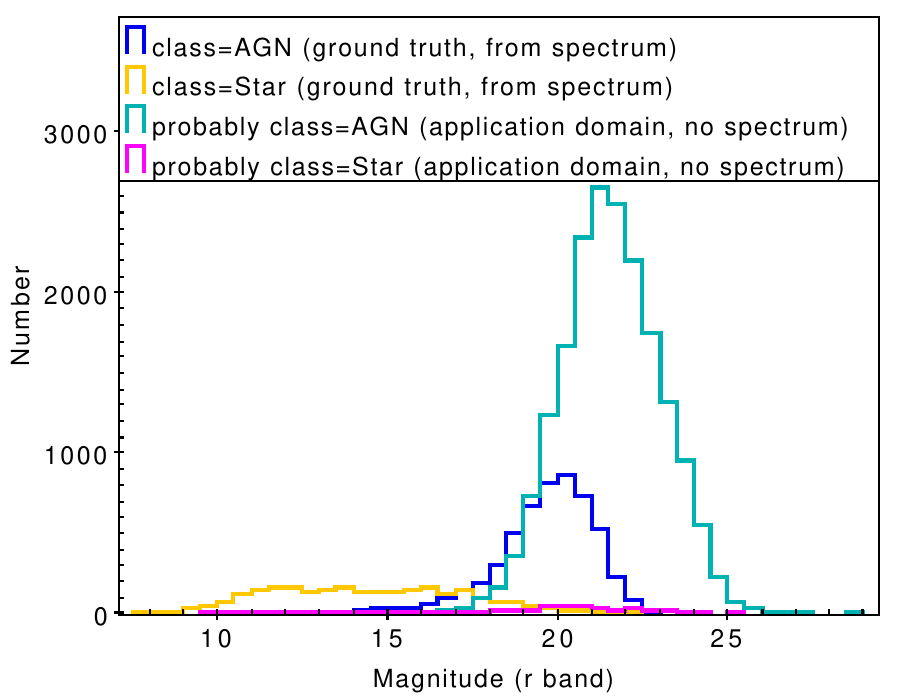}
    \caption{Illustration covariate shift from the eROSITA Full-depth equatorial survey (eFEDS; \cite{Salvato2022}). From expensive ground truth (spectroscopy), some of the sample was classified into AGN (blue) and stars (yellow), and can be used as a training sample for classification of the remaining samples. The distribution of one of the available and important input features, the r-band magnitude, is shown on the x-axis. The ground truth spectroscopy was obtained preferentially for bright sources, but is less available above magnitude 20. This means the application domain, where the classifier needs to work well, is substantially different to the training data. Similarly, for the large magenta distribution, photometric redshifts should be computed based on training on the (brighter) dark blue distribution. This is an example of covariate shift. The classes with ground truth are imbalanced, and the application domain shows a different class imbalance.
    }
    \label{fig:rmag}
\end{figure}
That machine learning \textit{learns} from a training sample, 
the training sample is a prior distribution. 
In regression, this can be readily seen when plotting the prediction 
against truth in the validation sample, where instead of a
1:1 line the predictions follow a sigmoid function, trending towards 
the bulk of the training sample distribution. This is illustrated in \ref{fig:regression-bias}.
Similarly, in classification
balanced training samples or imbalance corrections are needed.

The training sample is a set of elements in feature space, but by
itself, is insufficient to define a model (a function) 
that can predict from arbitrary points in the feature space.
In machine learning, the adopted model structure together with the 
stochastic training procedure imposes smoothness over these
samples. Model predictions are sensitive to the training sample
and adopted loss function.

\subsection{Vary your prior}

To get as many predictions right as possible, it seems 
natural to adopt a summarizing prediction error score.
However, one scientist may be interested in the right tail
of the distribution (for example, the reliability of predicted redshifts above 6),
and not care about the performance within the left tail.
Then, the summarizing score for training may need to be customized.

Such a scenario may apply to your project too. Consider building several
predictors, each optimized to a different region of the parameter
space or sub-tasks. Then, consider the one appropriate for the question.
The ensemble of predictions may also be a useful indicator of uncertainties 
due to model assumptions.

\subsection{Take care of covariate shift}

The training data and final application data set where the method
should make predictions will be slightly different. This is because
clean training labels are obtained typically for the brightest and
nearest astronomical sources.
So systematically, fainter and more distant sources will be missing in the training sample, but may be common in application.
This shift in the input features when going from the training set
to the application domain is called \emph{covariate shift}.

Figure~\ref{fig:rmag} illustrates this effect, which is pervasive in astronomy. The magnitude distribution of the data set where ground truth is available is skewed to lower values (brighter on the sky). These are the blue and yellow histograms. A useful machine learning classifier of stars vs AGN in this data set should however predict where ground truth is not available (magenta and purple histograms).

To address covariate shift, build a classifier (such as a Random Forest)
to distinguish training samples from final application samples. Compute
the probability for each sample to be in the training data, and use
that as a weight. An alternative is to divide the training data into
groups by probability score \cite{rosenbaum1984reducingbias}. 
Train the method on each group separately.
When applying, use the appropriate group. STACCATO \cite{Revsbech2018} goes
beyond this with training data augmentation appropriate for each group.
Also, weigh the performance measure on the test/validation sample
by the sample probability score. Otherwise, the real-world performance
may be mis-estimated.

For neural networks, another approach is domain adversarial training \cite{ganin2016domainadversarialtraining} (see \cite{Perdue2018advers} for an application in physics). There, in addition to the usual output objective, another classifier is trained simultaneously, and shares the same first few network layers, which act to extract features. The second classifier tries to distinguish data from the training sample and data from the application domain set.
During training, the feature extraction layers are trained to disallow this classifier to succeed. This prevents the final model from using information that would be biased towards the training data set.

\begin{figure}
\begin{centering}
\includegraphics[width=8cm]{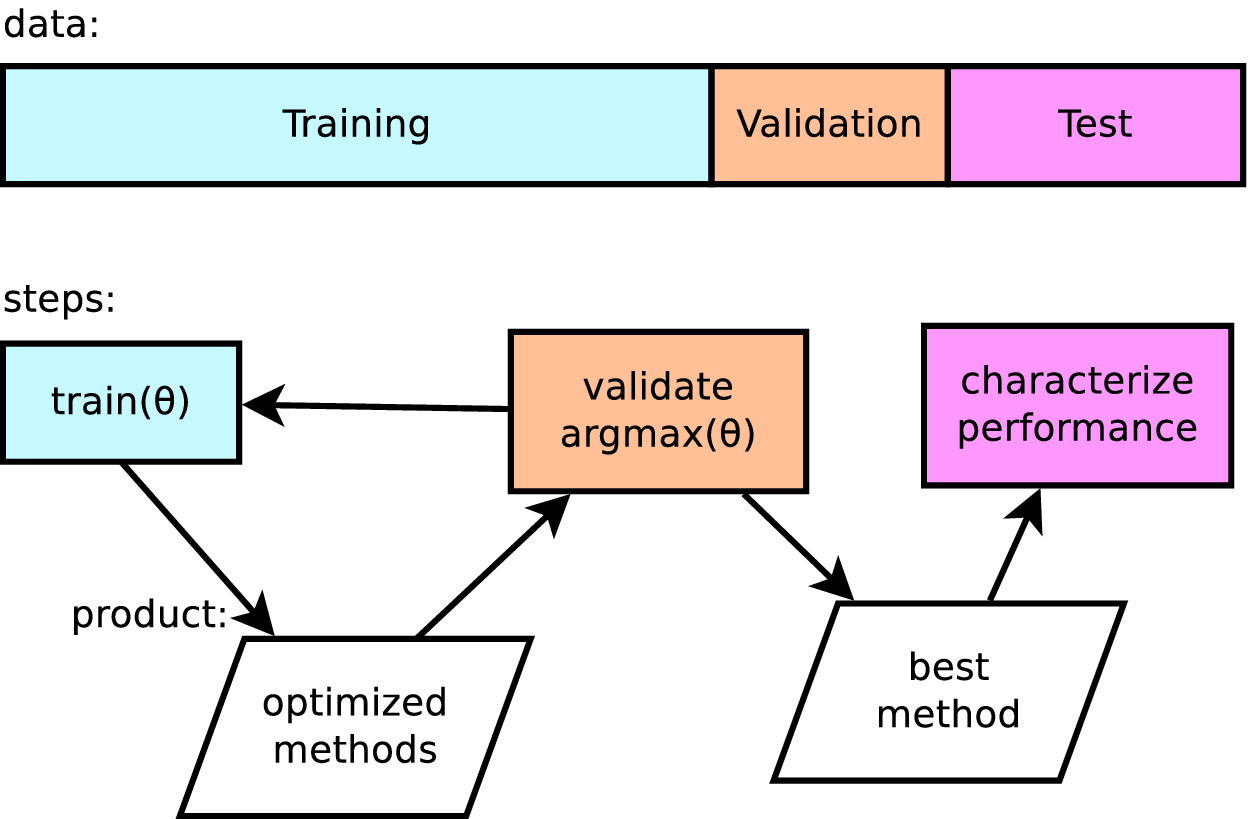}
\par\end{centering}
\caption{\label{fig:Typical-workflow}Typical workflow in introductory machine
learning tutorials and data challenges. 
The data set is split into a training set used to optimize hyper-parameters $\theta$, yielding a best method. The performance of this method is characterized with an unseen validation dataset. Some model improvements and iterations at this stage will lead to the final model. The test set, never seen by any model until this point, will give the final performance.
}
\end{figure}

\begin{figure}
\begin{centering}
\includegraphics[width=0.8\columnwidth]{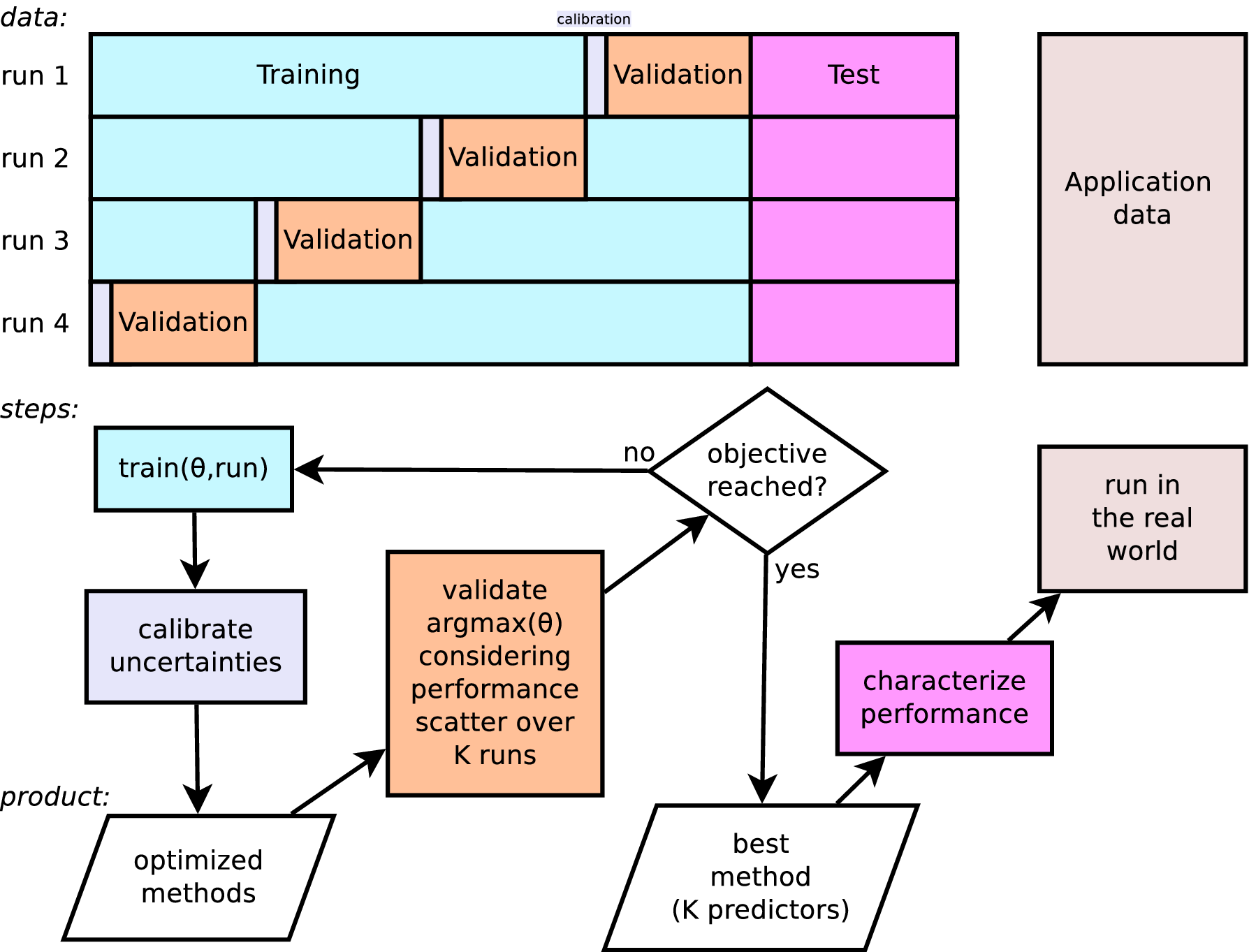}
\par\end{centering}
\caption{\label{fig:workflow-astro}A workflow for predicting with trustworthy
uncertainties. A small portion of the training data is held out to
calibrate the uncertainties. This can be considered part of the training.
Instead of training once, K-folding (here $K=4$) trains with different
seeds and data partitions. On the validation data set, the performance and
its variance are measured, and the best hyper-parameters $\theta$ are selected.
Among methods that are equally good within the error, we can use a 
secondary criterion like interpretability or speed to select one.
If the objective seems to be reached, 
the performance is finally characterized with the yet unseen test
data set. The model is delivered and applied to the real world data.}
\end{figure}

\subsection{One possible workflow}

A workflow typically found in machine learning tutorials is sketched
in figure~\ref{fig:Typical-workflow}. This includes (1) splitting
the data into training, validation, test samples, (2) training different methods and variants of a method with differing hyper-parameters (such as neural network architectures, optimization strategies) on the training data,
(3) choosing among these variants using the validation set, and (4)
finally assess the performance of the best method with the test
sample.

Regarding (1), a large enough test sample should be put aside, so that the performance indicators can be reliably computed, and that the entire domain is sampled. This can be done randomly, but for ordered data sets such as time series, one may want to predict from past to future and leave out the most recent observations. Complex data which exhibits a clustered structure should be resembled fairly in the test sample, to allow computing the performance indicators. If this is not done carefully, information can leak from the training sample into the test sample, a unfortunately common pitfall leading to over-estimates in model performance (see e.g., \cite{Kaufman2021leakage,kapoor_leakage_2023}).
Then, figure~\ref{fig:Typical-workflow} illustrates splitting the training and validation data sets, with 10-30 per cent remaining in the validation data set. The training data set should be as large as possible to allow the method to learn, but the validation data set needs to be large enough to be able to reliably distinguish competing models. Similar considerations on how to split (ordered, shuffled, etc.) apply.

The sections above suggest adding a few more steps, illustrated in
Figure~\ref{fig:workflow-astro}. This includes calibrating uncertainties
with a small sample removed from the training sample, 
K-fold cross-validation over
training and calibration to consider training variations and have
performance uncertainties (see sections~\ref{sec:insignificant}, \ref{sec:insignificant} and
\ref{sec:makeuncertainties}; nested cross-validation may be necessary).

The final step is the real-world application
to unlabeled data and having it be useful in this application domain. 
In astronomical instrumentation,
after commissioning and calibration on the mountain, the event of
getting real data from the sky is called \emph{first light}, and is
a celebrated achievement.

\section{Human learning}

Astronomy progresses by humans gaining insight into how the Universe
works. Scientific theories of how the universe could behave are falsified\cite{popper2005logic} by experiments, or remain standing.
For some machine learning problems, such as data cleaning and mere
discovery of objects (both essentially removing understood nuisances),
it may not be very important to understand how the machine does it. However,
for understanding cosmic populations, our limited detection of phenomena,
and the operational definition of what is being found by the machine,
it is important to look closer.

\subsection{What has the machine learned?}
\label{subsec:learning}

Astrophysics is familiar with another case where highly non-linear,
large-scale computing systems are to be understood. These are computer simulations that evolve fluids under gravity, heating, radiation, and/or magnetic fields. Such simulation boxes can produce complex phenomenology, including the formation of solar systems or galaxies. Each simulation snapshot may be gigabytes of raw data, which needs to be interpreted.
For example, we may want to understand a spatial grid where each cell encodes density with higher-level concepts, such as over-dense clumps, streams, and voids, their evolution, and the causes of this evolution. Below
are three techniques employed to learn from such complex simulations:

(1) One way science progresses is by a physics model failing to explain
data. In particular, when a physics model without an effect is discrepant,
and when the effect is added, the physics model explains the data, one could
infer that the effect is occurring in nature. This is covered under
\textbf{ablation studies} below.

(2) Another way to learn from a complex system is to simplify it down
to something that can be understood. The ideal example is to boil down a hydrodynamic model of some system to one formula. That analytic model
may not be as good or generic, but may capture the essential behavior of the complex model.
\textbf{Synthesizing the essence} into a \textit{proxy method} is how humans can learn from a machine,
and communicate its core idea.

(3) Finally, simulations can be tested on simplified systems
in isolation. For hydrodynamics, typical well-observed test examples\cite{Hopkins2015GIZMO} include blast waves or the mixing of two fluids of different densities.
Design such inputs to \textbf{interrogate the model} for what it thinks
should happen. Are there inputs where you can derive the output analytically?

These three approaches are expanded below, in turn.

\subsubsection{Ablation studies}

Let's say you have a deep neural network that works really well. It
processes 10 input features in addition to 4 groups of 1000 input
features.

To find out what the machine found useful, drop one input feature,
retrain, and see if the performance is comparable. It may be helpful
to drop a group of input features, or simplify the input space. For
example: replace colors with an average; drop every other column;
smooth the input. Another important technique is to force a symmetry
onto the input space \cite{Carleo2019}. This is problem dependent, but for example,
radially average each image to remove any angular information.

If the performance is still comparable, the information removed was
not very relevant, and you can resume with the simplified space. If
the performance has dropped substantially, you discovered an important
source of information that the model uses.

Taking away information is called an \textbf{\emph{ablation study}}.
There are other techniques for measuring feature importance, but ablation
studies have two benefits for astronomy: (1) if the input feature
is irrelevant, a costly measurement may be avoided, and (2) taking
away input information is easier to explain and understand than some 
other approaches.

\subsubsection{Ablation studies of the model}

It is easy to fool yourself with a nice flowery explanation of why
a method works. Remain skeptical of such explanations: The success
rates in predicting whether a method will work on another, easier seemingly
similar data set are very low. You will see experienced machine learning
people flat out say: ``We tried methods A, B and C. C happened to
work.'', without going further.

Indeed, it can be very difficult to judge whether one method works
while another does not. Beware of jumping to conclusions on the cause, 
without testing that effect directly. Related
to the ablation studies on the data, you can also take away part of
the model. For example, replace special sub-networks 
(e.g., capsule networks, attention networks), 
connections (e.g., skip connections in U-nets), 
activations (e.g., recurrent, gated, memory cells), learning algorithm
with a simpler or more common choice. This should demonstrate that
the more complex model component is necessary.
Such variations can also clear up the conceptions you
have built over the last few months about what was important to get
the method working, which may be confused by other changes introduced
simultaneously (on the data, performance quantification, variations
tested, etc). If such ablation studies reveal model simplifications 
with comparable performance, this may also lead to speed-ups.

\subsubsection{Interrogate the model}

For the interpretability and investigation of machine learning models, there is now an extensive literature. 
\cite{Zhang2021NNInterpretability} categorized these into extracting logic rules (discussed in section~\ref{subsec:learning}), interpretation by attribution and by hidden semantics (both discussed in section~\ref{subsubsec:blackbox}) and explanations by example (discussed below). 
In addition to the references in section~\ref{subsec:learning}, \cite{Fan2021NNInterpretability} provides an extensive survey of techniques.

One example is the design of adversarial examples. Can you find the smallest data change
that swaps the classification for the training samples\cite{goodfellow2014adversarialexamples} or creates an outlier? 
Discuss such cases within the team to determine whether they are problematic or acceptable.

What if X is different in the input data, how does the method react?
Set each input feature to the median value, except for one that is
varied. Plot the variation of the output as a function of this one
feature. This gives as many plots as input features. A further variation
is to set another feature to its 10\%, 50\% and 90\% range, by which
pair-wise sensitivity of the output can be investigated.

In some cases, we may have to accept that we cannot understand the 
model. We can still build confidence by characterizing its effective behaviour in detail. 
How well is it performing in each region of the feature space 
(e.g., quadrants in each panel of a pairs/corner plot)?
What are the main classes of outliers? 
Can potentially faulty outputs be recognized to caution the user?
Communicate these in publications and talks.

\subsection{Communicating}

Effective communication of machine learning models is critical for 
widespread adoption and further development.
In linear regression, papers can quote the model parameters in a small table,
and state the formulas. This allows colleagues to easily implement 
the model for their own work and build on it. For complex methods, the question arises
how the method will be accessible to other scientists. How will you
distribute the machine learning model? 
In a paper, how can you explain the method
in sufficient detail so the results can be reproduced? 
With how much effort?
Sharing open source code, cleaned training data, and the trained model, in a findable, accessible, interoperable and reusable (FAIR, \cite{Wilkinson2016Data}) way is recommended. 

A simpler method may be preferable, at least for some small tasks that are not the main subject of a publication. 
Simpler can be quantified by the number of configuration decisions to made, which also reflects in the length of text needed to describe all the steps necessary to reproduce the results on another person's computer.
This could be an additional criterion to choose between methods.

The bias assessment studies and extensive validation tests discussed in the previous sections are intended to understand a model's strengths, range of applicability, limitations and failure modes. To strengthen the scientific community's trust in the model, a publication should include these tests, their results and their interpretation.

In machine learning, much time and computing power is spent on trial
and error for finding techniques that work. It would be beneficial
if others can skip testing methods you already tried and did not work.
Such reports may ultimately lead to some wisdom on which
techniques work for which types of problems. Report (e.g., in an appendix)
variations that did not work at all, and hyper-parameter variations
that work equally well. Include as much as you think is helpful.

If the goal is to reach astronomers, then you should publish in astronomy journals. Helpful general guides for publishing papers in astronomy can be found in \cite{Chamba2022writingplanning,Knapen2022writingdeveloping} and 
\cite{Sterken2011EASWritingI,Sterken2011EASWritingII,Sterken2011EASWritingIII}. 

\subsection{Have stakes, make predictions}

In one project we discussed whether inputting ratios in addition to the individual numbers 
(photometry colors in addition to fluxes) into a deep neural network.
There were strong and heated arguments either way, and elaborate 
imaginations about what the deep learning model ``understands'' and can do.
To cut the discussion short, we decided to settle it with whether a ablation experiment shows a $>1\%$ change in performance. Making predictions and empirically testing them can lead to more cautious descriptions of deep neural network capabilities, and a deeper understanding.

If you think the model is good, you can also make a prediction on currently unlabeled data.
Can you predict the outcome of another experiment? This can be existing experiments
not involved in the training, or future follow-up experiments. These
should provide you with independent ground truth and check whether
the model has \emph{external validity}.
Even better if you can make a prediction in a space other than the
input training data space.

\section{Recommendations}

A machine-learning project should have a specific, measurable and useful objective, that can be validated with a test harness. 
At the start of the project, it is good practice to 
establish a set of trivial baselines and to define what a significant 
improvement over those would be. For scientific applications, ingesting and outputting uncertainties is expected. 
Reporting all trials and ablation studies contribute to a robust machine learning literature on which approaches work in which settings.

The final test of your project is to see it used in production. Good luck.

\bibliography{ml}

\section*{Acknowledgements}
The thoughts laid out here have been heavily influenced by conversations 
with colleagues and discussions in astronomy machine learning conferences such as EAS2022 in Valencia and ML-IAP2021 in Paris. If you have comments that should be included here, let us know. 

\section*{Author contributions}
The article was conceived, structured and initially written by JB. SF contributed to the writing in all aspects of the article.

\section*{Competing interests}
The authors declare no competing interests.




%


\end{document}